# Effects of COVID-19 Vaccine Developments and Rollout on the Capital Market - A Case Study

Maximilian Vierlboeck & Roshanak Rose Nilchiani, Ph.D.
*School of Systems and Enterprises, Stevens Institute of Technology, Hoboken, New Jersey, USA*

**Abstract** - Various companies have developed vaccines to combat the pandemic caused 2020 by the virus COVID-19. Such vaccines and the distribution can have a major impact on the success of pharmaceutical companies, which in turn can show itself in their valuation and stock price. This poses the question if and how the trends or popularity of the companies might be connected to the value and stock price of said entities. To gain some insight into these questions, the work at hand looks at five COVID vaccine development companies and evaluates their correlations over the development of the vaccine as well as after the rollout start. The process was conducted by using python including various libraries.

The result of this analysis was that there is a significant correlation between the Google Trend data and the respective stock prices (retrieved from yahoo! Finance) of the companies on average, where the time during the development of the drugs is more positively correlated and the post-rollout periods show a shift to a slightly negative inclining correlation. Furthermore, it was found that the smaller companies based on their market cap show a higher price volatility overall. In addition, higher average trend scores and thus popularity values were found after the rollout of the respective companies.

In conclusion, a correlations between the trend data and the financial values have been found and corroborate the plots of the data. Due to the small size of the sample, the result cannot yet be considered statistically significant, but possibility for expansion exists and is already being worked on.

## I. Introduction, Situation, & Methodology

The pandemic caused by the COVID-19 illness [1] has been dominating our lives for more than one year so far. Many people have been affected personally in various ways, such as financially, medical, and psychologically. Globally, millions of people have died from the virus and even today, over one year of the first death due to COVID-19, people are dying daily by the thousands.

Yet, there is light at the end of this dark tunnel since vaccines for the novel virus have been developed rapidly over the last year and various medications are now being administered. Various companies have developed vaccines and the following companies have so far planned or started the distribution: Moderna, Pfizer, Novavax, AstraZeneca, and Johnson & Johnson.

Not all of these companies have received approval for the distribution of their vaccines at the same time and some have been administered longer than others. Furthermore, the profit (or lack thereof) that these companies chose differs greatly, with some providing the vaccine not-for-profit until the end of the pandemic and some charing regular, self-determined prices. These differences and choices can have potential implications for the companies as a whole, which is why this paper set out to evaluate certain correlation and potential causations behind the vaccines, the development thereof, and the state of the companies at various times.

Unsurprisingly, the listed companies have been mentioned in the media spotlight on and off throughout the pandemic once their involvement and vaccine development became public and over time, all of the five companies were covered.

Given the aforementioned differences and the media coverage, the questions arises how the value and attractiveness of these companies changes and evolved over time during the pandemic. It might seem intuitive that there is a connection between the news and media coverage of these companies and their value, but this has to be evaluated. Thus, the work at hand analyzes the connection and or correlations between various factors and the values of the companies based on publicly accessible data that shall be combined in a model/framework to develop. Specifically, the following three questions are addressed by the research:
1. **Are the news related to vaccines impacting companies' values, if so, how?**
2. **How is the deployment of vaccines impacting companies' values?**
3. **How is the vaccine production affecting companies values and volatility?**

In order to address these questions and find potential correlations or even causations, the "Cross-Industry Standard Process for Data Mining" (CRISP) [2, 3] is being applied to develop and apply a python model that is capable of assessing the dynamics of the described companies. In CRISP, which focusses on extracting information from data, various phases are cycled through iteratively with repetitions where necessary and applicable. The six phases of the process are explained briefly hereinafter [2]:

1. <u>Business / Research - Understanding Phase:</u> This phase sets the objectives and requirements for the work and also outlines the data processing steps. In addition, a preliminary strategy is defined.
2. <u>Data Understanding Phase:</u> In this step, data is collected and explored to evaluate its quality assessed, which allows for the selection of appropriate sub-sets.
3. <u>Data Preparation Phase:</u> The preparation phase is used to shape the data into a usable format to allow for the subsequent processing.
4. <u>Modeling Phase:</u> This phase is to choose, calibrate, and apply proper modeling techniques to allow for evaluation of the results and derivation of insights.
5. <u>Evaluation Phase:</u> The second to last phase evaluates the application of the model and defines its applicability.
6. <u>Deployment Phase:</u> The last phase provides the actual application of the model and also includes preparation of the insights such as the creation of a report, for example.

These six phases form the foundation of the presented work and are processed successively. Hence, the paper at hand has been divided into five sections. In this first section, the first phase is conducted, which outlines the objectives and strategy for the work. The second section then outlines the data utilized and how it was being prepared for the processing in the model, thus covering phase 2 & 3. The following section then describes the model as well as its evaluation, corresponding to phase 4 & 5. The sixth phase is represented by the fourth section, in which the results are being shown and discussed. Lastly, the final section provides a conclusion and brief outlook how work might continue in the future. Figure 2 depicts the inclusion of the CRISP phases in the sections of the paper at hand.

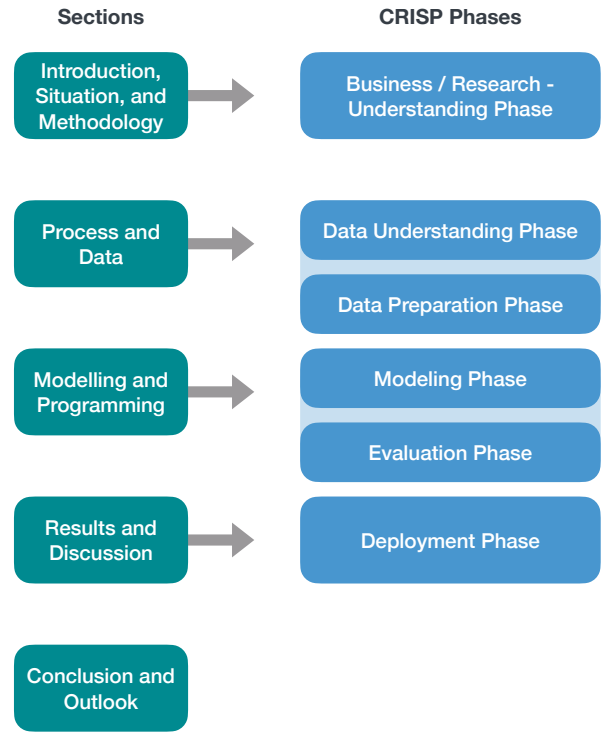

*Figure 2 – CRISP Implementation and Sections*

## II. LITERATURE REVIEW

In order to assess the situation and other research that has been conducted along the same lines as the topic presented, a literature review will be described below to give an overview for the general research of the topic act hand as wells the current state of the research including recent trends and developments.

When looking at market dynamics in general also considering specific influences, such as the ones this paper addresses, one can see that substantial work has been published. Topics published and researched cover market dynamics & macroeconomics [4-7], volatility and fluctuations [8-11], as well as behavioral and political influences [12-15]. It also has to be noted that the references researched go back far into the twentieth century when analysis approaches and assessments were less if at all supported by computational resources.

The described research above is also related to the attractiveness and potential gains of stock price predictions, which has also lead to other publications that outline approaches to predict the behavior of the stock market and specific companies with different methodologies, such as neural networks [16] and data analysis [17-20]. This also includes research that tries to connect specific events, such as product announcements, to the stock price of publicly traded companies [17].

More specifically in line with the presented research, recent publications show a growing interest in the behavior of the stock market in general as well as specific companies [21, 22] also in relation to the pandemic and its long lasting effects [23]. Since the unprecedented disturbances [24] and initial market crash caused by the pandemic not only impacted sectors directly related to the medical systems, for example, but even interfered with supply chains [25], a diverse and increased interest and research motivation can be seen in general. Some publications have already reached hundreds of citations in less than a year [26].

In addition, analyzing stock price and valuation with data analysis and even machine learning are trends that have recently emergent as well. For instance, Li et al. [27] showcase an attempt in which they combine financial information with data extracted form social media to evaluate what effect engagement such websites can have on stock prices for both, epidemic and non-epidemic circumstances. Other publications include assessments and implications of events and publicity regarding the tock price and thus valuation of companies [28, 29] and Zhu even provides a potential connection between vaccine incidents and stock price impacts for the pharmaceutical sector, although locally limited [29].

Furthermore, some very specific reseach publications were found that looked into the dynamics and effects behind vaccine developments and their relation to the stock market. For example, Alifa and Yunita [30] showed an increase in trading volume in the pharmaceutical sector caused by the announcement of COVID-19 vaccine trials by a major developer in Indonesia.

Similar to the presented research, Xia and Hu [31] assessed the connection and potential influences of investor attention (measured by the Bidu Index of certain pandemic related keywords), as outlined by Da, Engelberg, and Gao [32], on the stock market variables return rate, trading volume, and amplitude. The authors found a significant positive impact of investor attention on all three variables during the pandemic.

The literature review shows a substantial amount of reseach conducted in the field of the proposed topic as well as ongoing stream and trends with high attractiveness also due to the potential returns of the capital market enabled by new and predictive insights. The most recent publications show that the described work of this paper is in line with the current trends and even directly related research by Xia and Hu [31] was discovers. Yet, the aforementioned work differs from the presented one in various points, most importantly the data sources, the processing of the data, as well as the evaluated parameters. Xia and Hu [31] research the same connections though and thus, in addition to the other insights found by the literature review, it can be said that the research presented in this paper poses value and potential contributions to the current scientific trends.

With this overview at hand, the next section will outline the data utilized as well as the overall process applied and implemented.

## II. Process and Data

Since the analysis and work shown in this paper relies on data obtained from public sources, it also requires certain steps that make the data usable as it was not specifically provided nor created for the outlined purpose. Thus, the methodology described in this chapter was applied to obtain, prepare, and apply the respective datasets and data points.

In order to analyze the impact of the circumstances outlined in the three questions above, for each of the companies, the following data was collected:
- Historical prices and trade information from the website of yahoo! finance [33]
- Global trend data for each company [34] from the website of Google Trends [32]

The data from the described websites was obtained through the libraries and packages outlined in the next section. The timeframe for each dataset was set to begin on January 1st of 2020 and ended with the last available day (May 20, 2021 for the presented work). The exact processing and implementation of the data is described in detail in the next section which outlines the model and code developed.

As for the method to answer the questions set forth in the beginning, a correlation analysis supported by plots was chosen to asses if the trend and therefore popularity of each company has an impact and effect on the value, represented by the stock price. This analysis allows for a mathematical evaluation and also enables the assessment of each company on an individual basis before bringing them together for a final conclusion regrading the correlations overall and thus answers to the questions. The approach is depicted in Figure 3 below:

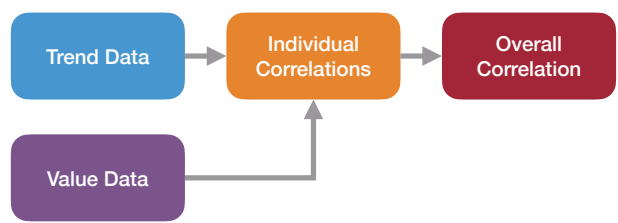

*Figure 3 – Analysis Foundation and Correlation*

Since the questions defined in the introduction are threefold, the analysis to conduct had to provide results that allow for the address of each question individually. Therefore, the following three assessments were derived to evaluate each of the three questions respectively while also allowing an overall assessment and resulting conclusion that shows potential overarching phenomena and possible dynamics.

1. <u>General</u> correlation between company stock price and news trends.
2. Correlation between the stock price and the trends of the company before the first deployment of the vaccine (<u>pre-rollout</u>).
3. Correlation between the stock price and the trends of the company after the first deployment of the vaccine (<u>post-rollout</u>).

Since the data obtained from the sources was not immediately usable, it was necessary to organize and prepare it for further processing. The processes used therefore is described hereinafter.

With the libraries described in the next section, the data could be obtained in the form of pandas [35] frames/structures. In order to weed out the unnecessary information, the columns were reduced to only include the necessary values. From the Google Trends, the following information was retained: the date of the record, the scale used by the library to adjust the value to the absolute, the actual trend value. From the yahoo! Finance data, the following values were kept: the date of the records, the prices at Open, Close, High, and Low.

The cleaned structures were then exported for both datasets (each company separately) as comma separated value (csv) files. This allowed for skipping of the mining in the future and also enabled backups and process tracing. Such backups were also created after the described merging, which resulted in three files per company: trend, value, and merged data (see Figure 4).

The inclusion of dates allowed for a simplified merging of the two datasets by just adding them together based on the date column. This way, another cleaning step could be implemented without specific coding, which is the removal of holidays, weekends, and other miscellaneous days where no trading, and therefore no value changes, occurred. The merged and remaining data was thus ready for further processing.

Since it is possible that some of the above described steps might have introduced errors into the dataset, it was essential to verify the data before further processing. Therefore, each of the merged files was manually cross-checked at random spots in the dataset for consistency and correctness based on the online sources. No errors were found and this verification was repeated at different dates to ensure the correctness of the processing outcome with different datasets.

The last piece of data missing for the correlation analysis were the specific rollout dates of each company. Said dates were researched online and gathered based on news articles. The following dates outlined in Table 1 were implemented. For Novavax, the rollout has not started as of May 11, 2021 and hence, no date could be considered.

| Company | Date |
| --- | --- |
| Moderna | 12/21/2020 |
| Pfizer | 12/14/2020 |
| NovaVax | N/A |
| Astra Zeneca | 1/4/2021 |
| Johnson & Johnson | 3/2/2021 |

*Table 1 – Vaccine Rollout Start Dates [36–39]*

Finally, with the prepared and verified data, the analysis was conducted with the results obtained from the process above. This evaluation included the comparison and averaging of correlation factors to allow for a final verdict and answers to the three main questions. By including the results of each company, the dynamics of the industry and overall circumstances could be taken into account instead of just relying on the behavior of one company that might have been affected by various other influencing factors. The exclusion and mitigation of the influence of such factors is especially important when it comes to volatile parameters such as the valuation of a business and or its stock price, as a plethora of factors affect the behavior of these aspect, such as management decisions, scandals, or earnings, for example.

Overall, the described methodology and utilized data allowed for a comprehensive look at various companies and thus the deduction of potential correlations affecting all of them. The exact implementation of the described steps is outlined in the next section, which includes the exact description of the python model and code procedure developed to address the questions defined.

Lastly, it has to be noted that all data and code that was written and utilized can be provided upon request for reference purposes and also for expansion and utilization within the frame of other research work and projects.

## III. MODELLING AND PROGRAMMING

As previously mentioned, the language chosen for the work described was python and for the code outlined hereinafter, the version 3.8 as well as the IDE of PyCharm [40] was utilized. Furthermore, the following libraries and packages were used due to the listed reasons:

- pandas [35]: pandas data frames was used to process tabular and conduct calculations, such as the Pearson's Correlation Coefficients [41].
- pytrends [42]: this library was used to obtain the datasets and make then accessible in the form of pandas data frames.
- yfinance [43]: the library of yfinance was utilized to retrieve the stock price data of the respective companies in form of pandas data frames.
- matplotlib [44]: this library was used to create plots and other visualizations of different kinds for easy and consistent depiction of the acquired data.

The acquisition and cleaning/merging of the data was already outlined in the previous section, which is why this section focuses on and explain the actual concatenation of steps behind the code. All code was written independently and no resources made by others were used besides the libraries indicated above and their respective documentation/samples.

The first step conducted with the cleaned and merged data was the derivation of the scatter plots. These plots were created for the total time period, the pre-rollout, and the post-rollout phase. For this operation, the dates outlined in Table 1 were used in conjunction with the matplotlib library. This allowed for the creation of the plots analyzed in the next section. Since the creation of the plots was repetitive, this operation was conducted in various loops to create all outputs.

After the creation of the scatter plots, the correlation analysis was conducted. This analysis was addressed by the correlation function native to the pandas library. This allowed for the efficient extraction of the correlation matrixes that compare all factors with one another. Thus, the results shown in Figure 9 in the next section were obtained for each company for the respective time frames of the entire period, pre, and post-rollout. This allowed for the overall analysis that enabled the discussion and answering of the questions set in the introduction. For the sake of completeness, Figure 6 shows the entire process from start to finish with the respective results and information obtained in each step. In addition, the used libraries and loops are indicated with the respective symbols after the data preparation.

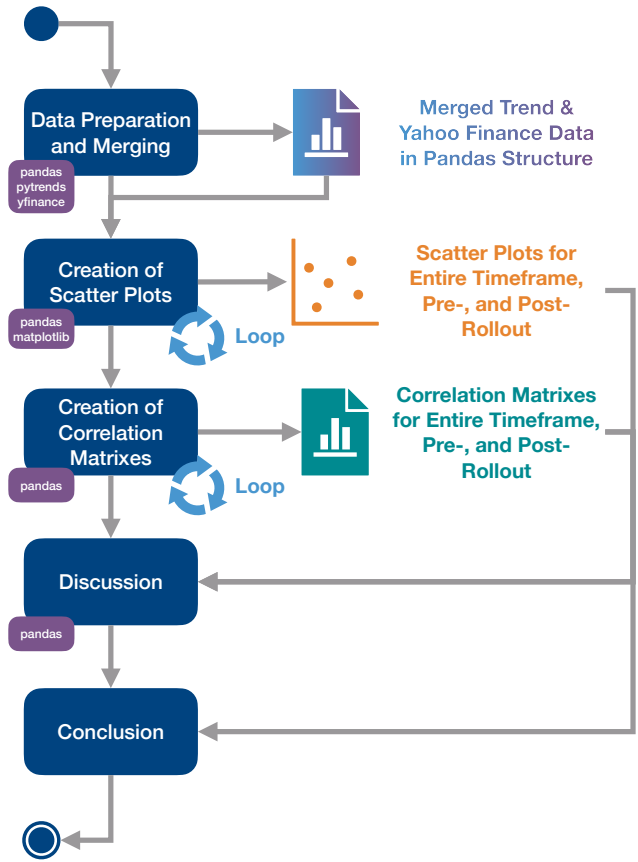

*Figure 6 – Complete Model of the Developed Process*

In order to verify and thus define the applicability of the created model, the results of the code were cross checked at certain points via debugging to ensure the correct calculations and results. In addition, all datasets were used with certain check variables (such as the below described scale value) to quickly discern the validity and potential errors in the mathematical applications.

With the created outputs, the results could be discussed which is described in the following section.

## IV. RESULTS AND DISCUSSION

The results obtained from the data and mathematical processing are twofold. Scatter plots were created to show the overall data and after that, the correlation of the various was generated as Correlation matrixes.

The first result, the scatter plots, are depicted in Figures 7 & 8 and discussed afterwards. In said figures, the trend value is depicted as the score relative to the absolute maximum trend recorded during the time frame (1/1/20 through 4/13/2021) and the closing price is depicted in USD. It has to be noted that the plots were not significantly different when another price was used compared to the closing price.

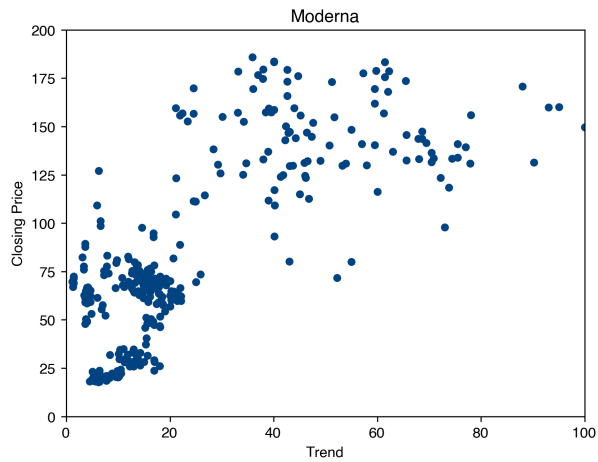
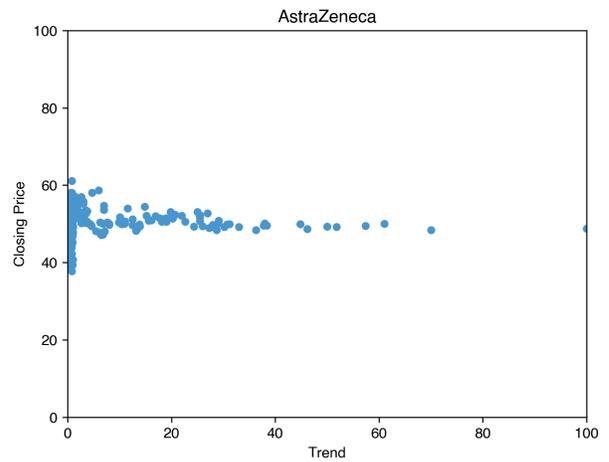
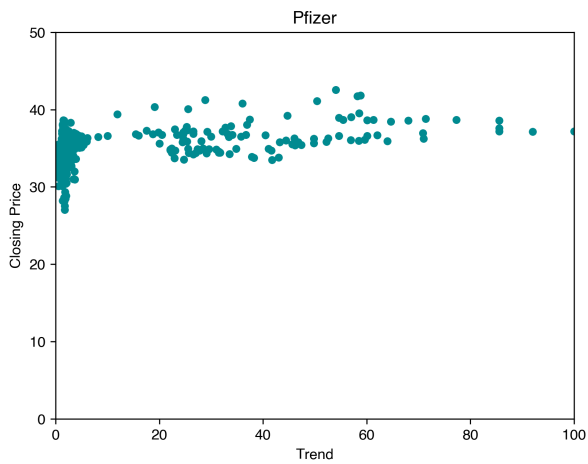
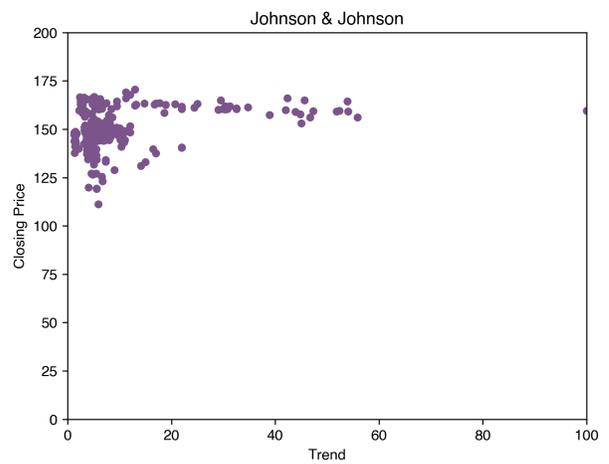
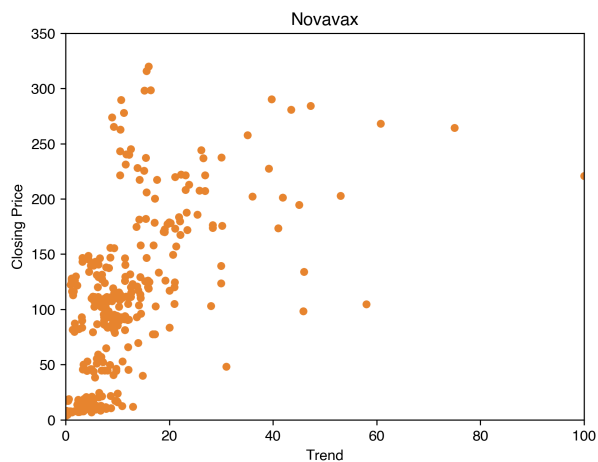

*Figure 7 – Company Value/Trend Scatter Plots for Closing Price – Complete Time Frame*

Since the periods post and pre-rollout were also of interest due to the different question in the introduction, the different time periods and respective scatterplots are depicted in Figure for all companies except Novavax, for which no vaccine release date has been set yet. For the plots in Figure 8, the scale and axes were kept the same compared to Figure 7 to allow for a clear visual comparison.

In Figure 8, the pre-rollout scatter plots are on the left and the post-rollout plots respectively on the right. The axis range depict the entire span in all cases for the x axis and for the y axis, the entire range covering the maximum price is depicted to allow for a comparison regarding vertical spread and volatility.

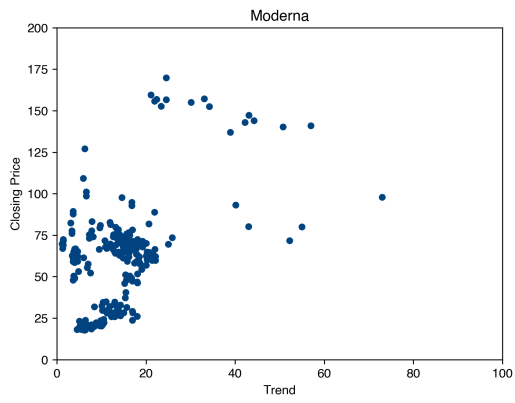
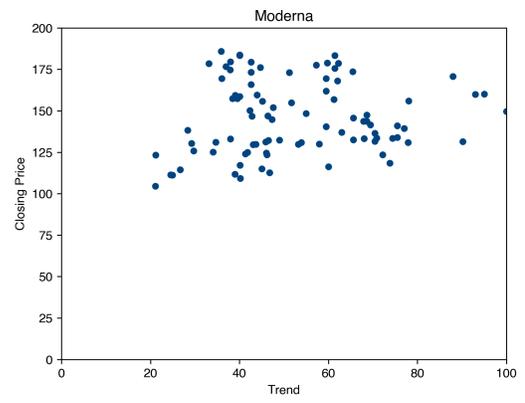
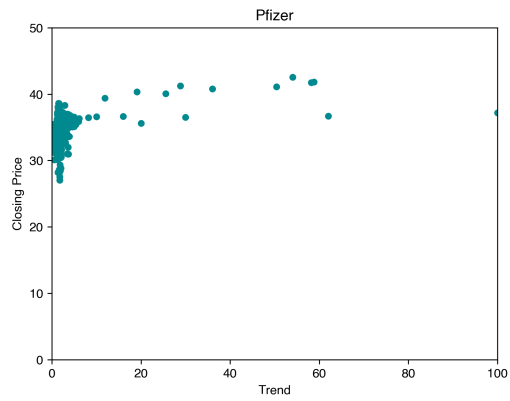
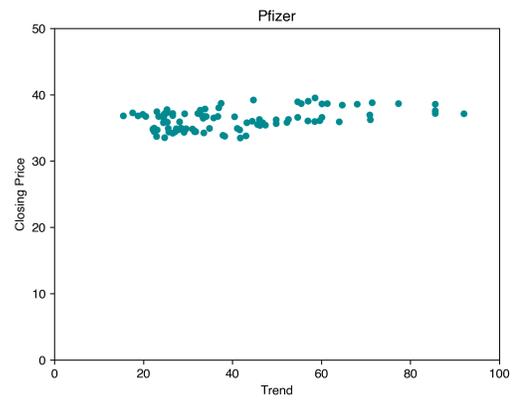
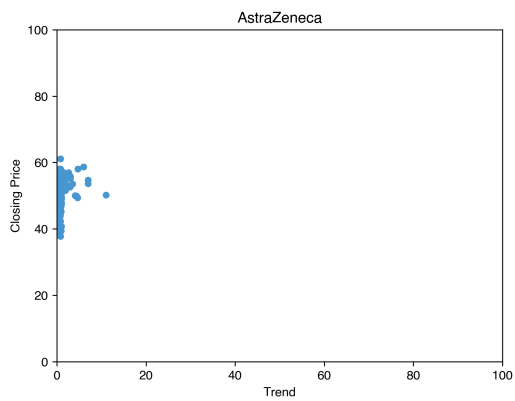
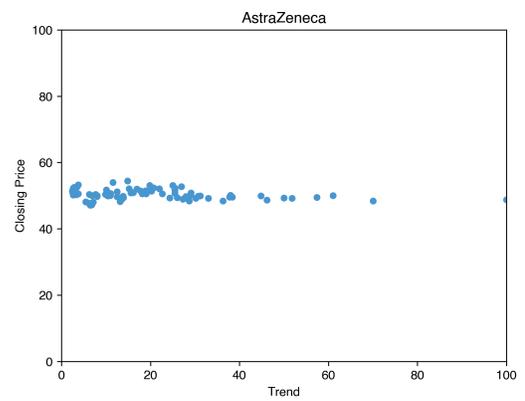
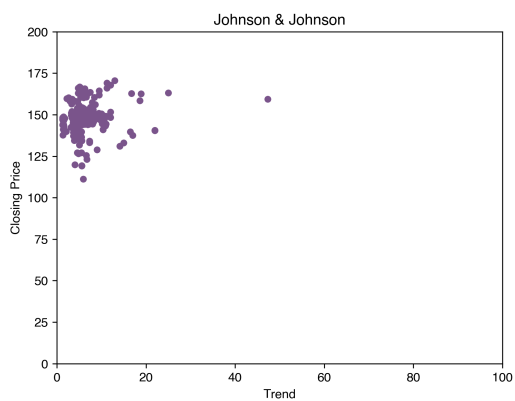
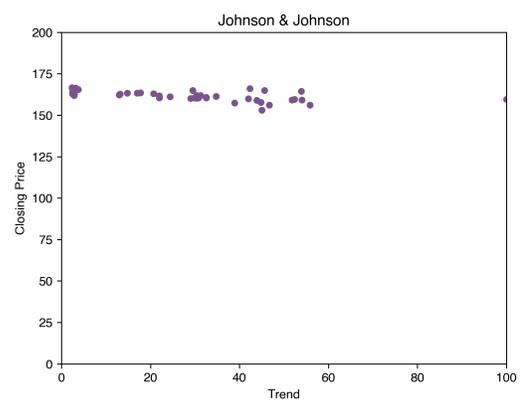

*Figure 8 – Company Value/Trend Scatter Plots for Closing Price – pre-rollout period left, post-rollout right*

The scatter plots already allow for some important insights, which shall be outlined in the following paragraphs.

First, it can be seen that some companies showed a much higher value fluctuation than others. For example, Moderna's closing price over the assessed time period ranged from under $25 to over $175, whereas Pfizer's staid consistently within the $30-40 range.

Second, the trend data differs greatly between the pre- and post-rollout periods. For all of the assessed companies, the trend data before the start of the rollout was predominantly in the lower regions close to zero. After the respective rollouts though, little to no 0 values are seen and the overall volatility increased way into the higher areas of the trend value. For all companies except Pfizer, the maximum also happened post-rollout.

Third, the plots post-rollout show on average higher values for the closing price and the minimum stock price for all companies lies before the respective rollouts. This could be simply due to the fact that the market has been recovering and the stock prices therefore rise in general, which results in higher average prices later in time, but it is also possible that the rollout is seen favorable in general by investors, which yields higher valuations.

These insights gained by the scatter plots already provide some partial answers to the questions in the introduction, but cannot be the sole foundation for a conclusion, thus, the remaining paragraphs of this section discusses correlation in a quantitative way.

For the evaluation of the correlation, the Pearson's Correlation Coefficient [41] (PCC) was utilized. This coefficient can be calculated directly from the pandas structures and was thus evaluated for the time same time periods as the scatter plots for all prices included in the datasets.

It has to be noted that the scale factor, which indicates the reference of the trend data, has been kept in the correlation matrixes as a control value. While it does not add any information to the correlation between value of the companies and their trend/popularity, it serves as a check to see if the data was processes correctly. Since the scale is directly related to the trend value, it should at all times show a close, if not identical correlation to all other values and a very high correlation to the trend column. Thus, if the correlation $c_{sx}$ between scale $s$ and any other variable $x$ would have differed greatly from the correlation $c_{tx}$ of the trend $t$ with the variable $x$, an error in the calculation, processing, or data could heave been indicated. No such anomaly was found in the results.

For the calculated Pearson's Correlation Coefficient, values above an absolute of 0.6 indicate a strong correlation (with 1 being perfect), values between 0.3 and 0.6 indicate a moderate correlation, and factors between 0 and 0.3 are considered weak [45, 46]. Thus, the color coding in the subsequent tables indicates the type of correlation according to the derived Pearson correlation factor. Green indicates positive correlations ( > 0.3), orange indicates insignificant correlations ($-0.3 < c < 0.3$), and red indicates negative correlations ( < − 0.3). The saturation of the color depicts the significance/strength of the correlation.

| Company | Open | Close | High | Low | Avrg. |
|---|---|---|---|---|---|
| Moderna | 0.7661 | 0.7657 | 0.7669 | 0.7669 | 0.7664 |
| Pfizer | 0.4462 | 0.4432 | 0.4525 | 0.4421 | 0.4460 |
| NovaVax | 0.5770 | 0.5892 | 0.5921 | 0.5698 | 0.5820 |
| Astra Zeneca | -0.1280 | -0.1431 | -0.1483 | -0.1201 | -0.1349 |
| Johnson & Johnson | 0.3174 | 0.3284 | 0.3387 | 0.3150 | 0.3249 |
| Average | 0.3957 | 0.3967 | 0.4004 | 0.3947 | 0.3969 |

*Table 2 - Trend Correlation Data - Complete Time Frame*

The first insight gained from the correlation table, which applied to all three, not just Table 2, is the fact that the different price that can be looked at, i.e. opening, closing, high, or low prices, do not have any visible influence on the correlation. These values show similar factors for all companies with only slight and deviations.

More importantly, the table for the entire timeframe shows an overall positive correlation with the exception of AstraZeneca, whose correlation is negative, but almost insignificant (absolute at or under 0.1 [46]) according to the Pearson's Coefficient.

In addition, we also see that the Moderna company shows the strongest correlation between stock price and trend data, almost reaching a correlation factor of 0.75 on average. This indicates that overall, there is a significant correlation and positive link between trend/popularity and value/stock price. This indicates that the stock prices rise with higher popularity on average.

Yet, since the table for the entire time frame is only a snapshot, the respective time periods before and after the rollout begin are important as well and were evaluated based on Tables 3 and 4 below. Furthermore, the combinatory aspect and influence of all time frames has to be considered, also due to variances in weight/relevance that can result in different clusters.

| Company | Open | Close | High | Low | Avrg. |
|---|---|---|---|---|---|
| Moderna | 0.5543 | 0.5217 | 0.5484 | 0.5205 | 0.5362 |
| Pfizer | 0.4671 | 0.4174 | 0.4889 | 0.3923 | 0.4414 |
| NovaVax | N/A | N/A | N/A | N/A | N/A |
| Astra Zeneca | 0.1200 | 0.0850 | 0.1197 | 0.2934 | 0.1545 |
| Johnson & Johnson | 0.1589 | 0.1545 | 0.1710 | 0.1487 | 0.1583 |
| Average | 0.3251 | 0.2946 | 0.3320 | 0.3387 | 0.3226 |

*Table 3 - Trend Correlation Data Pre-Rollout*

Table 2, which shows the data for the correlations pre-rollout, already depicts a different picture compared to the entire time period. The correlations are significantly lower and overall, the average correlations are thus almost insignificant. This indicates that the time frame before the rollout is not as strong overall, which could be due to the high fluctuations or lower causalities. In addition, the separation of the time periods can influence the correlation as well, as smaller datasets might show different correlations within due to their respective positions on the scatter plots, for example (see discussion at the end of this section).

| Company | Open | Close | High | Low | Avrg. |
|---|---|---|---|---|---|
| Moderna | 0.0604 | 0.1134 | 0.0710 | 0.0111 | 0.0640 |
| Pfizer | 0.3615 | 0.4133 | 0.3788 | 0.4079 | 0.3904 |
| NovaVax | N/A | N/A | N/A | N/A | N/A |
| Astra Zeneca | -0.2418 | -0.2519 | -0.2612 | -0.2299 | -0.2462 |
| Johnson & Johnson | -0.6433 | -0.5797 | -0.6086 | -0.6081 | -0.6099 |
| Average | -0.1158 | -0.0762 | -0.1050 | -0.1047 | -0.1004 |

*Table 4 - Trend Correlation Data Post-Rollout*

Table 4, showing the correlation data post-rollout, shows that after the rollout, no positive correlation can be found for any of the companies. On the contrary, some of the companies show negative correlations that are to be considered moderate and thus, with higher trends/popularity, their stock price declines. This can be due to the fact that after the rollout, the news we have seen so far are predominantly problems with the distribution, which are factors that can negatively influence the stock price. The fact that the rollouts all happened recently within the last few months might amplify this effect, as we can see with Johnson & Johnson for example, since the company recently had their vaccine distribution paused relatively shortly after distribution begin [47].

In total, it can be deduce that some correlations can be seen in the data evaluated and that it seems like overall the value of companies might show a slight correlation with the trends and their popularity. Yet, the data is not consistent and does fluctuate over time specifically influenced by the rollout as it seems. This event seems to have a significant impact and the correlation factors decrease significantly for all companies after the rollout, which turns most of them negative, even for the most positive ones pre-rollout. This phenomenon could be related to the fact that news/trends before the rollout are mostly anticipation and therefore positive, whereas news and popularity after the rollout could be related mostly to mistakes and other unexpected occurrences, which can be mostly considered unpleasant. Also, the shift seen is probably amplified by the fact that the interest in the vaccine and companies changes post-rollout. While before the rollout, interest of investors might be high, after the rollout, personal interests, such as side effects might take over and cause the shift in correlation. Such phenomenas could be evaluated by using sentiment analysis or specific text processing in the future to deduce the exact reasons for such shifts possibly even with the inclusion of machine learning.

Another factor to consider is the profit stance of the different companies when it comes to the vaccine. Moderna and Pfizer have made their vaccines for-profit, while Johnson & Johnson and AstraZeneca have decided to provide their drug for free until the pandemic ends [48]. This is in line with the highest correlation factors seen for Moderna and Pfizer overall, which seems to indicate that investors might have higher expectations and thus bigger investments in the for-profit companies.

Overall, the correlations seen in the tables corroborate what the scatter plots have shown. The different behaviors shown pre- and post-rollout are seen in the scatter plots as well as the tables and indicate different dynamics for those time periods. Yet, the size of these sub-dataset is something that has to be considered when comparing the dynamics as the dataset post-rollout is significantly smaller compared to the one before the distribution. Therefore, the behavior shown after the start of the respective rollouts can be influenced by its small size and thus small dynamics might be over-amplified. It is very well possible that the behavior seen after the rollouts are blips and would show much smaller effects if continued over a long time.

Lastly, the size of the sample has to be discussed to define and assess the statistic significance of the results. Since the number of companies was limited due to the nature of the topic, a sample size of ice (5) cannot be seen as significant and therefore, more data is necessary to actually provide statistically sound results.

With the outcome of the discussions and the results now at hand, the last section can take a look at the conclusion, answer the initial questions, and also give a brief outlook how work could continue.

## V. Conclusion and Outlook

The research at hand took a look at the business of the COVID-19 vaccines and involved companies to evaluate if and how the trending and popularity of such companies correlates with their values and stock prices. In order to address these hypotheses, a python process was setup to evaluate scatter plots and correlation matrixes for all companies and the timeframes before and after the rollout, as well as the entire period from 1/1/2020 through 4/13/2021.

After the evaluation and result discussion, it can be said that the conducted analysis and processing has shown valuable insight and allows now for the final answering of the questions set forth in the introduction.

1. **Are the news related to vaccines impacting companies' values, if so, how?**
   Based on the trend data, the popularities show an overall positive correlation with the value/stock prices of the companies regardless of the price type used for the evaluation. Moderna, Pfizer, Novavax, and Johnson & Johnson showed a significant and strong positive correlation over the entire time frame. AstraZeneca showed no significant correlation. In addition, the companies with the lower market cap (Moderna and Novavax) showed significantly higher volatility of the stock price regardless of the time period considered.

2. **How is the deployment of vaccines impacting companies' values?**
   Prior to the vaccine rollout of each company, which is assumed to be proportional to the effect of the development, the correlation is slightly reduced compared to the overall time frame, but still significantly positive. Thus, the conclusion can be made that for the assessed companies, there is a positive correlation on average between the popularity/trends of the companies during the vaccine development.

3. **How is the vaccine production affecting companies values and volatility?**
   The correlations show a shift when assessed for the time period after the rollout when vaccines are produced and distributed. For all companies, the correlation significantly declined after the distribution began, which allows for the conclusion that the rollout start poses a shift in the connection and correlation. For all companies, the correlation after the distribution start is either negligible or significantly negative, which indicates a different type of popularity potentially based on side effects or distribution problems.

Furthermore, the insights and results of the presented work are in line with other publications as described in the Literature review and show similar phenomena compared to what Xia and Hu [31] discovered for different circumstances and parameters, for example. This adds support to the validity of the work at hand and also further strengthens the potential of future extension and continuation of the presented approach.

It has to be noted that the sample size of five (5) companies cannot be regarded as statistically significant yet and that the size of the data set as well as the size ratio fo the data points before and after the rollout can have introduced some distortion or misleading amplifications.

These limitations lead to the future possible work as the developed framework can be extended to allow for a more comprehensive and thus potentially statistically significant case study. For example, including distributors or other involved companies might increase the number of included data sets and thus further support the validity of the findings.

In addition, several updates to the code basis can be made in order to increase its efficiency and or reduce the reliance on external sources, for example on manually retrieved dates for important events, such as the distribution begin. Also, expansions to get different insights such as the above mentioned sentiment analysis or specific text processing, and even natural language processing with potential machine learning can be implemented to expand the capabilities of the framework.

All in all, the conducted analysis and developed framework show some important insights and answered the questions set forth in the introduction satisfactorily. With the outlined potential expansion points for future work, the developed framework and code basis show promising possibilities to evolve into a comprehensive and useful application case.